\documentclass[fleqn,usenatbib]{mnras}

\usepackage{newtxtext,newtxmath}
\usepackage[T1]{fontenc}
\usepackage{ae,aecompl}

\usepackage{graphicx}	% Including figure files
\usepackage{amsmath}	% Advanced maths commands
\usepackage{amssymb}	% Extra maths symbols
\usepackage{natbib}
\usepackage{subfigure}
\usepackage{ textcomp }
\usepackage{soul}
\usepackage{lineno}
\title[The  variations of polarization in {S}5 0716+714]{The intraday variations of polarization vector direction in radio source {S}5 0716+714 }

\author[E. S. Shablovinskaya et al.]{
Elena S. Shablovinskaya,$^{1}$\thanks{E-mail: e.shablie@yandex.com}
Viktor L. Afanasiev$^{1}$
\\
$^{1}$Special Astrophysical Observatory of Russian Academy of Science, Nizhnii Arkhyz, 369167 Russia
}

\date{Accepted XXX. Received YYY; in original form ZZZ}

\pubyear{2018}

\begin{document}
\label{firstpage}
\pagerange{\pageref{firstpage}--\pageref{lastpage}}
\maketitle

\begin{abstract}
The bright radio source {S}5 0716+714, that is usually classified as BL Lac object, {is one} of the most intensively studied. {S}5 0716+714 demonstrates extremely peculiar properties such as the shortest time-scale of optical and polarimetric variations observed in blazars. In given paper{,} the results of 9-hour polarimetric monitoring of {S}5 0716+714 with the $\sim$70-second {resolution} carried out by the 6-m telescope BTA of SAO RAS, are presented. The observation data analysis {reveals} {the  variability} both in total and polarized light on the 1-1.5-hour time-scale. Since polarization is generated by plasma motion in {the} magnetic field, the variations of polarization vector are bounded with {the} magnetic field configuration and the average time-scales specify the size of {the} emitting region, unresolved directly. We suggest a processed numerically model of {the} polarization in {the} jet with {the} precessing helical magnetic field. Fitting the model {discovers} that observed short-term variations with {the} complicated trajectory of polarization vector could be induced by {the} magnetic field precession with the period $T \approx 15$ days.
\end{abstract}


%%%%%%%%%%%%%%%%%%%%%%%%%%%%%%%%%%%%%%%%%%%%%%%%%%

%%%%%%%%%%%%%%%%% BODY OF PAPER %%%%%%%%%%%%%%%%%%

\section{Introduction}

The object {S5} 0716+714 (R.A.: $07^h21^m53^s.4$, Dec.: $+71^{\circ}20'36''$) is a bright radio source with extraordinary variable brightness. According to \citet{Biermann1981}, {S5} 0716+714 was classified as BL Lac object due to specific features common to the class: flat {power-law} spectrum ($\alpha \geq -0.5, \ S_{\nu} \propto \nu^{\alpha}$ in {radio band}), high and variable optical polarization degree and significant variability in all wavebands. However, some authors mention fluctuations of intensity {shown} by {S5} 0716+714 -- up to 5 mag in flares \citep{Larionov2013} and up to 0.5 mag within the night -- are irrelevant to the sample of BL Lac objects.

{Though} the investigation of {S5} 0716+714 in all wavebands has been conducted for {over} 25 years, the object redshift is {still} an open {issue}. There {is} no sign of host-galaxy neither by direct photometric nor by spectral observations: there are {no} spectral features detected at the 0.3\% level in optical band \citep{Nilsson2008}. The {\it HST} survey of BL Lacertae objects \citep{Urry2000} had revealed {no} evidence of the host existence and therefore the severe restriction to the {maximum} host-galaxy brightness was given -- $m>20.0$ mag {that corresponded to the} redshift $z>0.5$. However, the {most} authors {accepts} $z=0.1 \div 0.5$, {obtained} by ground-based and still not concerted observations. The problem of {S5} 0716+714 redshift presently is not solved, and hence the extragalactic origin is not granted{.} 

The {S5} 0716+714 brightness is variable across all the spectrum from radio -- on the time-scale of days \citep{Liu2017}, $8-16$ hours \citep{Gorshkov2011a, Gorshkov2011b} -- to $\gamma$  -- $15$ min \citep{Gupta2009}. According to the general model \citep{Gabuzda2013}, the optical radiation of blazar is produced in {the} unresolved region of the jet at the distance $\lesssim 10^{-2} $ pc from the centre ({so-called} optical jet),  {and} the investigation of {the regions} {near} the central black hole {is} essential {for such issues as jet formation and variability origin}. Inside the jet{,} relativistic plasma radiates due to the motion in {the} ordered magnetic field producing non-thermal (synchrotron) polarized emission. The direction of polarization is determined by charged particles oscillations and so by their direction in the magnetic field. Therefore, the change of {the} polarization vector is a pointer of the field configuration.

{At} the optical jet scale plasma moves rapidly in a compact region that leads to polarization vector variability on short time-scales (intraday variability). The rapid variability both in total and polarized light was discovered: in  \citep{Impey2000} the significant polarization changes within less than $2$ hours were pointed out; the similar result was obtained in  \citep{Amirkhanyan2006} -- within $\sim$3 hours of {the} observation set polarization degree noticeably changed with the amplitude of 6\%. {More recent papers} \citep[e.g.][]{Zhang2018} {revealed the intraday variability as well.}

In {the} given work{,} the results of {the} {rapid variability investigation} in total and polarized light of {S5} 0716+714 are presented.  The obtained {observational data} {are} interpreted {within the} model of {the} precessing helical magnetic field. Finally, {the} constructed numerical model qualitatively explains our observations {and}  other authors' data. 

\section{Observations}

On February 2018 our group conducted a polarimetric monitoring of {S5} 0716+714 to reveal intraday variability of flux and polarization. The observations were carried out with the SCORPIO focal reducer \citep{AfAm2012} at the primary focus of the 6-m BTA telescope of the Special Astrophysical Observatory of the Russian Academy of Sciences (SAO RAS) {for} 9 hours. We obtained 460 exposures each of $60$ sec in {the} g-SDSS filter, interrupted by approximately $10$ sec gaps. 

A double Wollaston prism  \citep{Geyer1995,Oliva1996} was used as a polarization analyser. The {key benefit} of observations with double Wollaston prism is that measurements in mutually perpendicular directions $0^{\circ}$, $90^{\circ}$ and $45^{\circ}$, $135^{\circ}$ are obtained jointly: { the frame consists of four different images of the exit pupil, so the registration of the linear polarization parameters -- the Stokes parameters $Q$ and $U$ -- is simultaneous.} This is particularly important {for} unstable atmospheric conditions: as the parameters of the radiative transfer in the atmosphere vary, {scintillations and depolarization coefficients variations occur}. Concerning the observations in SAO, the main reason of depolarization is non-selective aerosol light scattering, {so} polarization changes up to several percents on time-scales of 10-30 msec. 

\begin{figure}
\centering
\includegraphics[width=0.75\columnwidth]{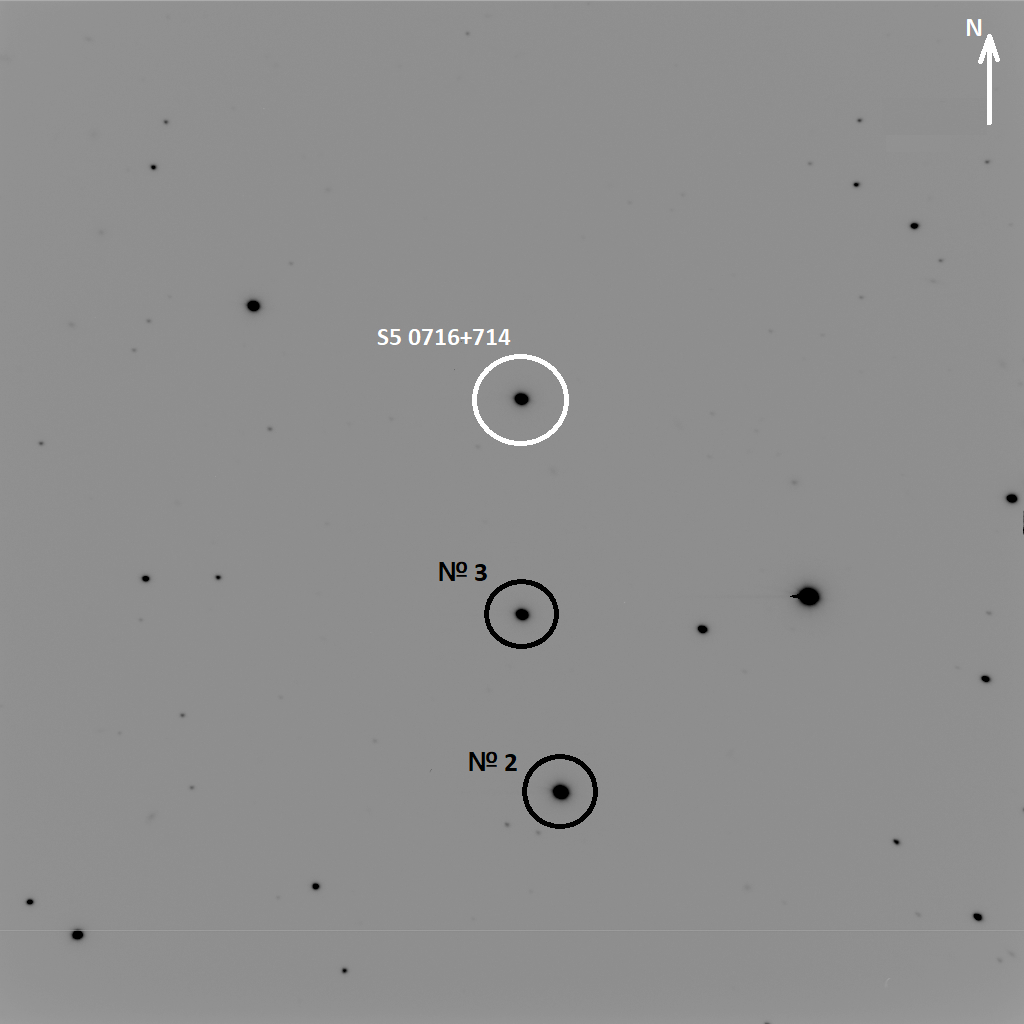}
\caption{The finding chart of {S5} 0716+714 {(FoV - $6.5' \times 6.5')$}.}
\label{map}
\end{figure}

{The analyser was rotated to keep in the field of view not only S5 0716+714 but also two stars close to the object} -- in Fig. \ref{map} \textnumero 3 (R.A.: $07^h21^m54^s.36$, DEC.:$+71^{\circ}19'20''.92$,  $m_B = 14.15$) and \textnumero 2 (R.A.: $07^h21^m52^s.33$, DEC.:$+71^{\circ}18'17''.71$,  $m_B = 13.04$). The first star was used as both a photometric and zero polarization standard \citep{Amirkhanyan2006};  { the star \textnumero 2 was used to control accuracy}. Then simultaneous observations of the object and the standard {admit} to correct unsteady atmospheric absorption and, moreover, to minimize depolarization. The normalized Stokes parameters $Q$ and $U$ {could be derived from the relations}:
    $$ 
    \frac{Q}{I} = \frac{I_0 - I_{90} D_Q }{I_0 + I_{90} D_Q} , \   
    \frac{U}{I} = \frac{I_{45} - I_{135} D_U}{I_{45} + I_{135} D_U} ,
    $$
where $D_Q$ and $D_U$ are coefficients of polarization channel transmission, calculated from {the deviation of the independently measured standard star. $D_Q$ and $D_U$ are weakly time-dependent, so considered to be constant: $D_Q = 1.036 \pm 0.015$ and $D_U = 0.985 \pm 0.015$}. The total flux is defined:
    $$
    I = I_{0} + I_{90} D_Q + I_{45} + I_{135} D_U.
    $$

\begin{figure}
\centering
\includegraphics[width=0.73\columnwidth,angle =90]{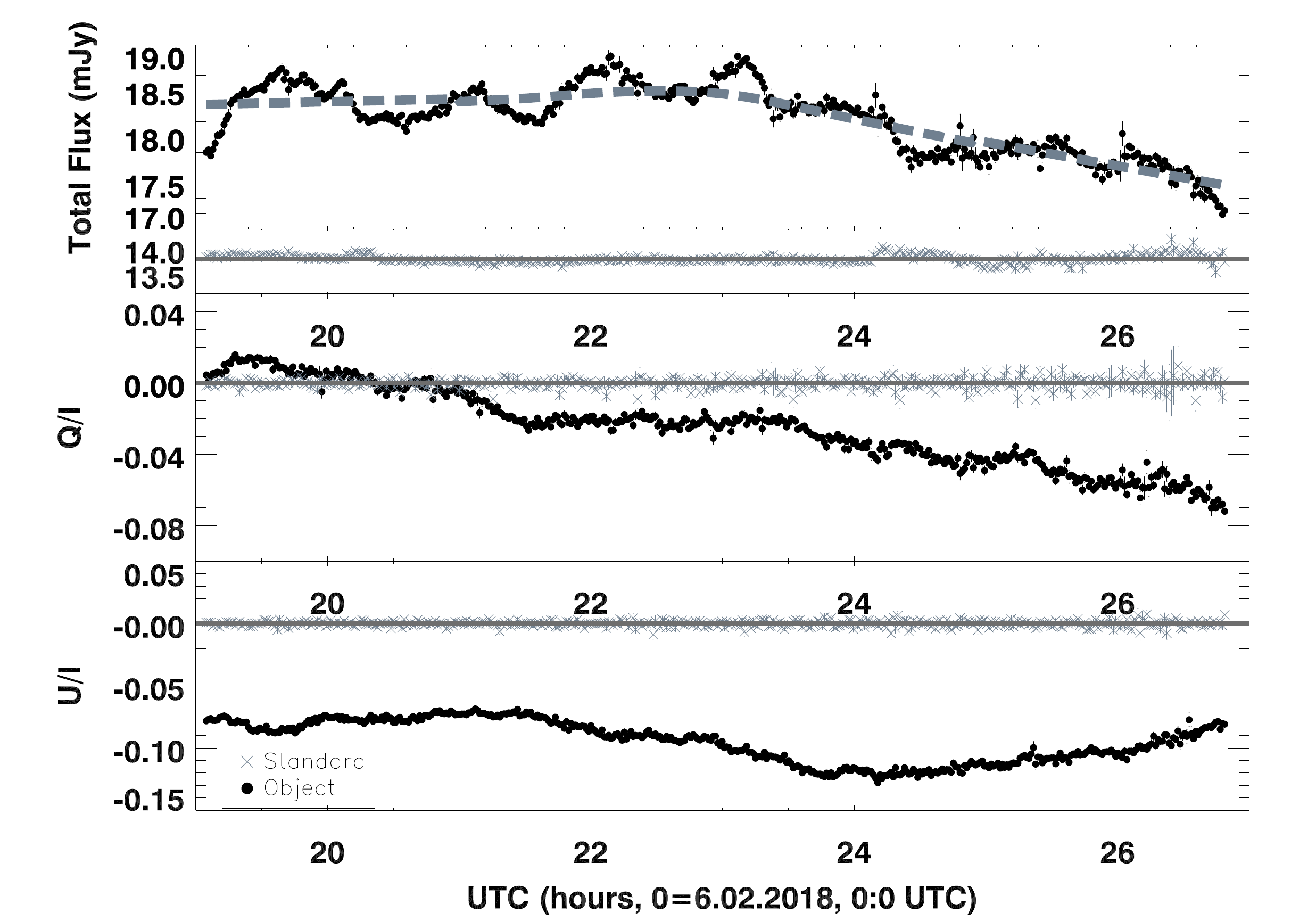}
\caption{The variations of the total flux and the Stokes parameters $Q$ and $U$ within the night. {The observations were started at 19:08 on 2nd February 2018 (UTC).}}
\label{FQU}
\end{figure}

The proposed {measurement} technique {provides} the photometry accuracy above $0.005$ mag and the {polarimetry} accuracy up to 0.1\% {on} average. The results of our observations {could be found} in Fig. \ref{FQU}.

\subsection{The flux variations} 

The variations of {S}5 0716+714 total flux are presented {in} the upper panel of Fig. \ref{FQU} with black dots. The standard star flux is shown with grey crosses for comparison. The standard deviation {of the flux} is  $\sigma = 0.079$. The zenith distance of the object was changing from $27^{\circ}.6$ to $55^{\circ}.2$. {While} the comparison star {has been} observed at the same height{,} {the effects of changing atmospheric absorption cancel out}.

\begin{figure}
\centering
\includegraphics[width=0.83\columnwidth,angle =90]{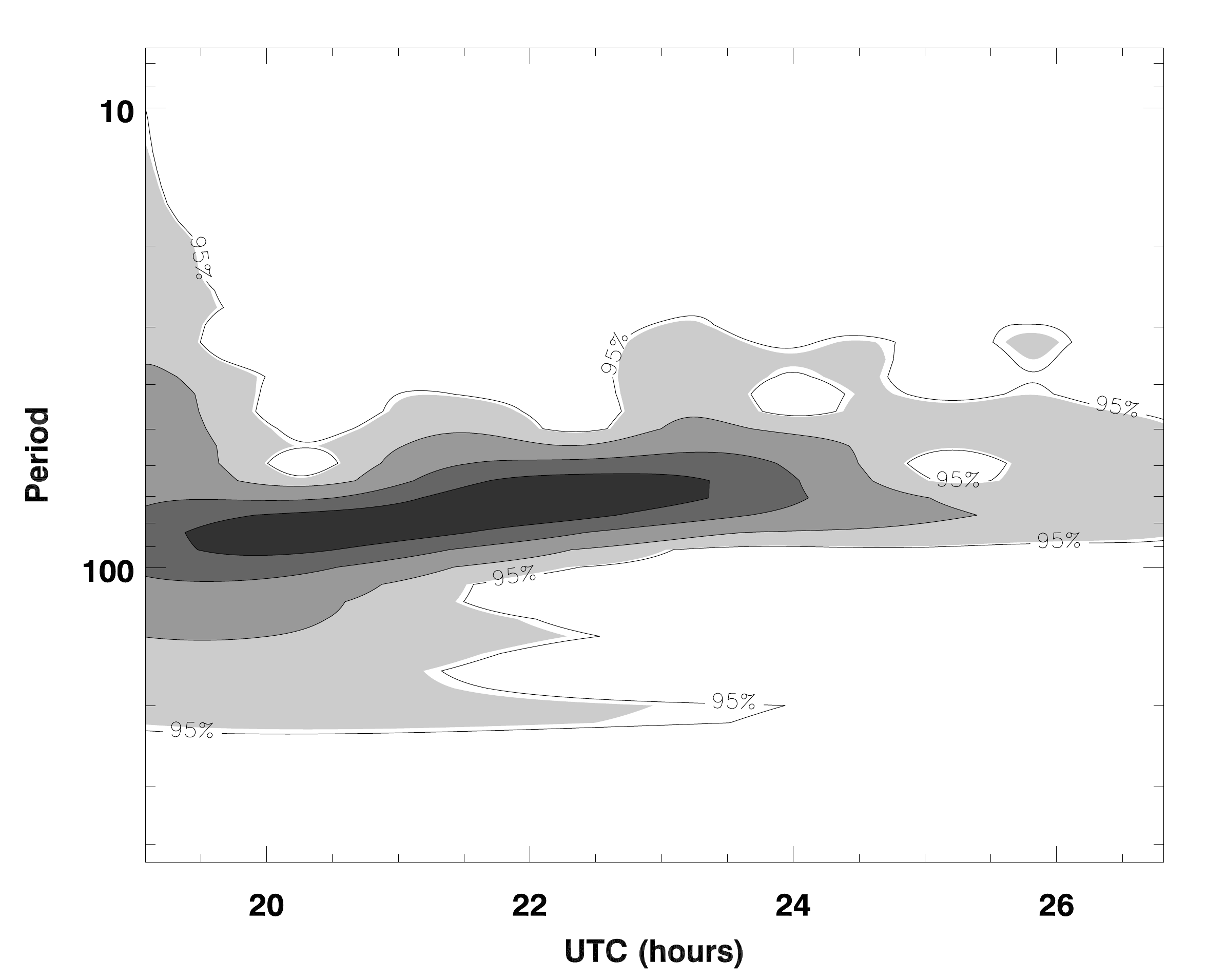}
\caption{The magnitude of the wavelet transform.}
\label{wv}
\end{figure}

\begin{figure}
\centering
\includegraphics[width=0.83\columnwidth,angle =90]{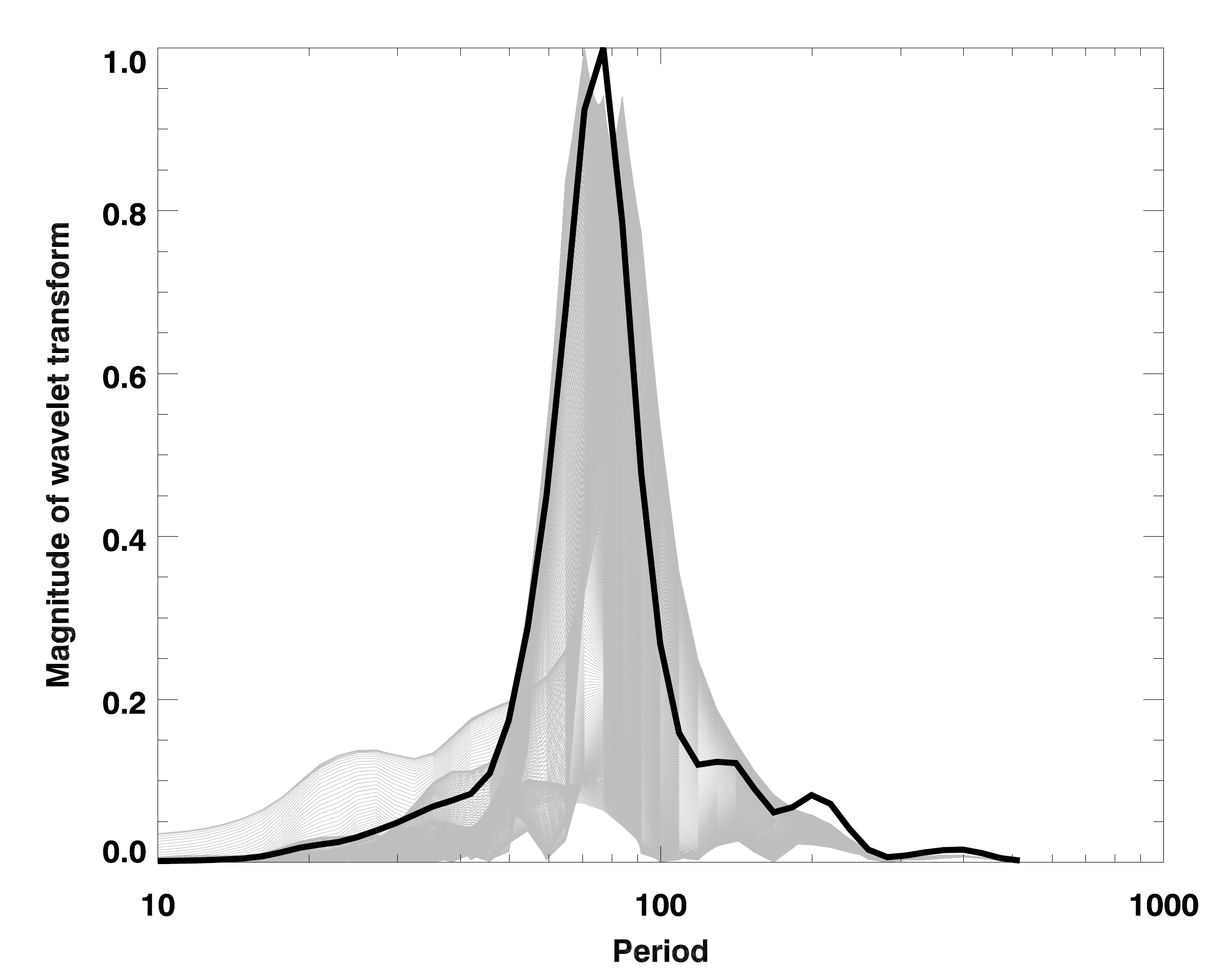}
\caption{The wavelet transform profiles (grey) and their median value (black).}
\label{wv_prof}
\end{figure}

Also, {throughout the night the total flux demonstrates the slight slope -- the trend, and the quasi-periodic variations, not detected for the standard star. In Fig.} \ref{FQU} {the trend is shown with a grey dash line.  To investigate the periodicity of the source flux variations with the subtracted trend the wavelet analysis is applied to follow the dynamics of the harmonic component. The magnitude of the wavelet transform is presented in Fig.} \ref{wv}, {where the areas of $0.7 \sigma$, $1 \sigma$, $1.5 \sigma$ and $2 \sigma$ (or 95\%) deviation are countered. Evidently, the variation period {is changing} from 65 to 85 min within the night. Median wavelet magnitude profile (Fig.} \ref{wv_prof}) {reveals the average period -- $\sim$77$\pm$10 min. }

\subsection{The polarization variations}

\begin{figure}
\centering
\includegraphics[width=0.83\columnwidth,angle =90]{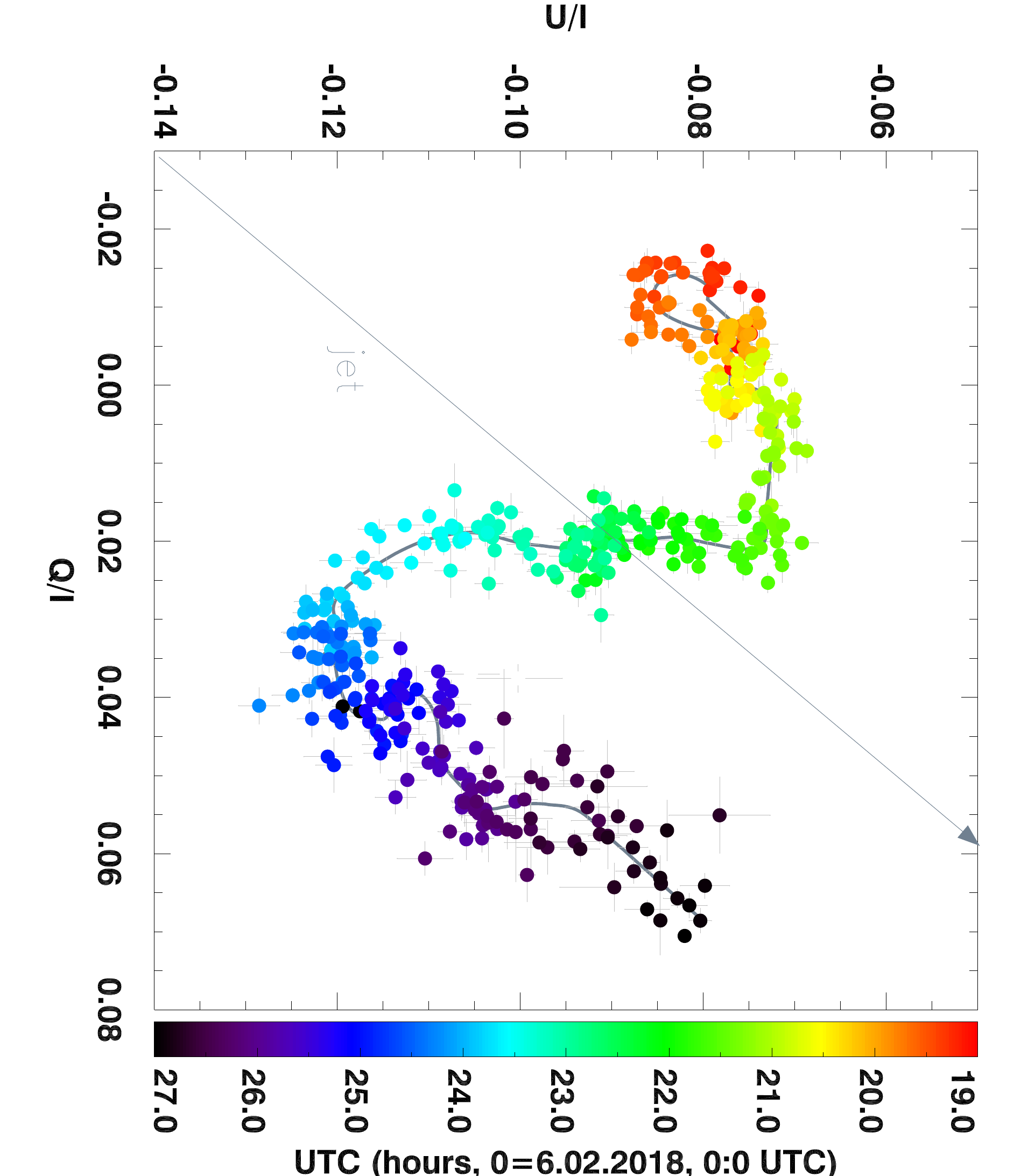}
\caption{The variations of normalized Stokes parameters $Q$ and $U$ within the night on the $QU$-diagram.}
\label{dragon}
\end{figure}

The variations of linear polarization -- the normalized Stokes parameters $Q$ and $U$ -- are shown {in} the middle and bottom panels of Fig. \ref{FQU}, respectively. {Similarly to the flux}, the {S5} 0716+714 {polarimetric} data are presented with black dots, the standard star data -- with grey crosses. The standard {deviations are} $\sigma_Q = 0.0038$ and $\sigma_U = 0.0023$. {Evidently}, within the night{, the} polarization {has} significantly changed  {by as much as} 7\%, yet more substantial inferences could not be done. {As the vector quantity, the polarization variations are performed on the $QU$-diagram as on the most convenient pictorial rendition}. Fig. \ref{dragon} demonstrates the motion of {the} polarization vector during the observation set; the {colours} correspond to time. {Moving along some complicated path, the vector switches the direction of the motion with the average time $\sim$1.5-3 hours. It should be noticed that polarization variations time-scale is of the order for the total flux ones. The physical interpretation would be given in Section } \ref{disc}.

To examine the polarization vector {variability} together with {the} jet location {the optical jet is assumed to coincide} with the {radio jet} at the position angle $\sim 25^{\circ}$ \citep{Rastorgueva2011}. As on the $QU$-plane the angles double, the jet position is $50^{\circ}$ (see Fig. \ref{dragon}). Then it occurs that the fluctuations of {the} polarization vector and, consequentially, the charged particles motion are perpendicular to the jet direction.

{An important point} is that in several papers \citep[e.g.][]{Impey2000} the variations of flux and polarization on the time-scales of 1-10 min are asserted found out. Actually{,} it could be suggested from the Fig. \ref{FQU} and \ref{dragon}, yet comparing with the standard {one could reveal} this kind of scintillations {is} the atmospheric noises, not connected with the source.  

\section{Model of polarization in jet}

The application of {magnetohydrodynamic} (MHD) models to astrophysical jets \citep[e.g.][]{Meier2001} {reveals} that inside the flow the instabilities such as kink-instability could develop that leads to {a} helical structure formation in plasma \citep[e.g.][]{Zhang2017}. The observations of the near active galaxies such as M87 \citep{Capetti1997} {confirm} this theoretical model. In {the} given section{,} we would like to {suggest} our geometrical model of polarization in {the} jet with the helical structure of {the} magnetic field.

\begin{figure}
\centering
\includegraphics[scale=0.6]{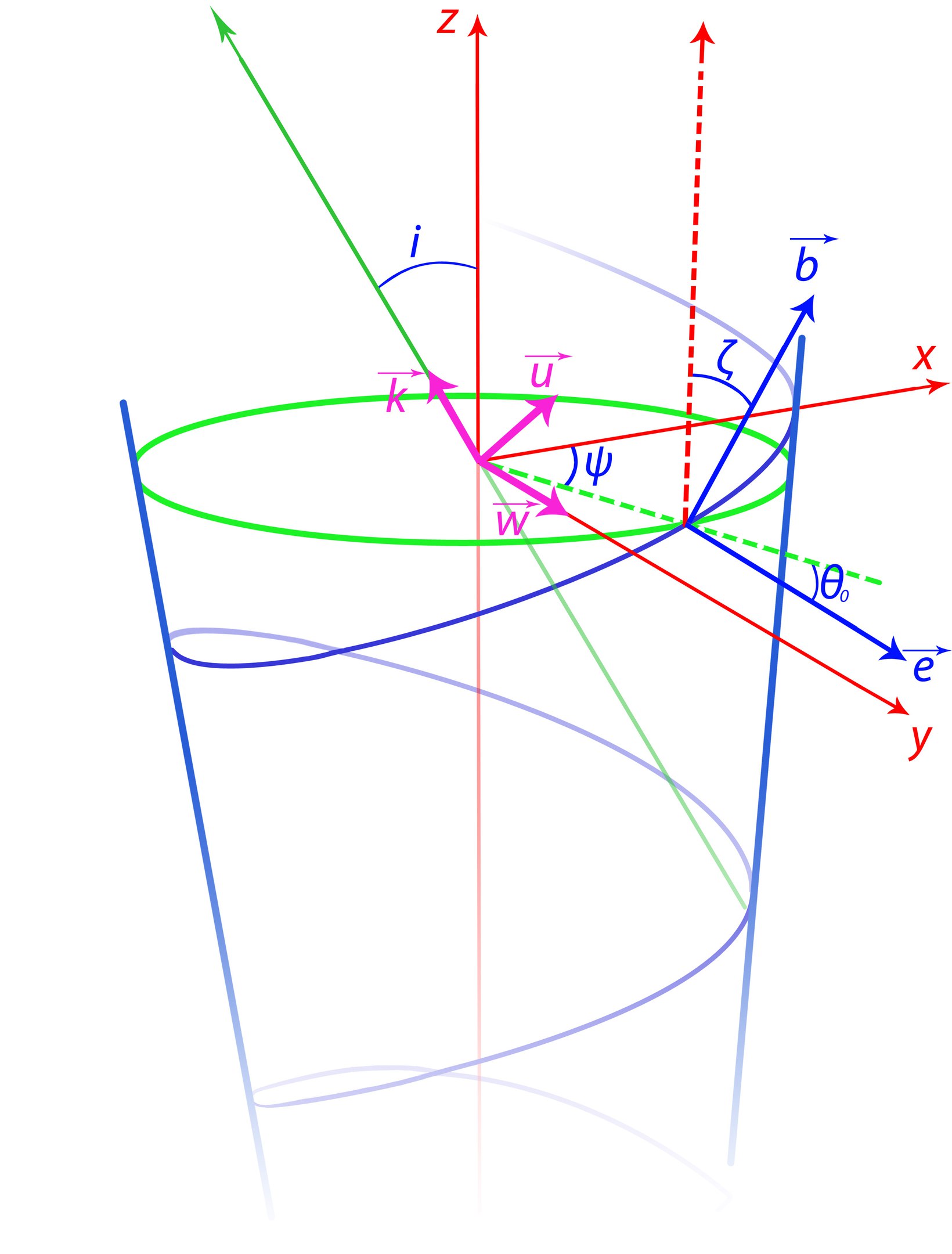}
\caption{The reference systems established to describe charged particle motion in jet.}
\label{sys}
\end{figure}

{To describe the jet geometry {a} reference system $(x, y, z)$ is applied, for which $x$-direction corresponds to the jet projection {to} the celestial plane for the observer, $z$ is directed along the jet, and $y$ forms an orthogonal basis with $x$ and $z$ } (Fig. \ref{sys}). The plasma motion in {the} helical magnetic field in {the} conical jet is described in {a} cylindrical frame $(\rho, \psi, z) $ where $\rho$ is the distance from the point to {the} applicate axis, $ \psi $ is {the} azimuth and $z$ is {the distance along the applicate} coincided with $z$-axis in $(x,y,z)$. $\psi$ is counted from $x$-direction. Assuming angular momentum $L$, kinematic energy $E_{kin}$ and half-opening angle $\theta$ to be constant, the equations of motion become \citep{Steffan1995, Li2018}:
$$
\rho = f \sqrt{  1 + \left( \frac{at+b}{f} \right)^2  } , \ \ \ \dot{\rho} = \frac{a}{\rho} (at + b)
$$
$$
\psi = \frac{\left[ arctg \left( \frac{at+b}{f} \right) - arctg \left( \frac{b}{f} \right) \right]}{sin \theta} , \ \ \dot{ \psi} = \frac{a f}{\rho^2 sin \theta} 
$$
$$
z=\frac{\rho - \rho_0}{tg \theta}, \dot{z}=\frac{\dot{\rho}}{tg \theta},
$$
where the next constants are used:
$$
a=\beta \sin \theta, \ \ b=\sqrt{\rho_0^2 - f^2}, \ \ f=\frac{j}{\beta},
$$
where $\beta = v/c$ is the physical speed of a jet component in units of the speed of light $c$, $\rho_0$ is the cylindrical distance, $j=\frac{L}{E_{kin}} c$ is angular momentum in units of distance. Time $t$ is also expressed in distance units, i.e. $t \equiv t c$.

Fig. \ref{sys} {displays that instantaneous velocity of the charged particle would be characterized by the vector ${\bf  b}$ along the particle direction and the vector ${\bf  e}$ perpendicular to the cone side and helical trajectory as well. ${\bf  e}$ indicates electric vector, indeed, therefore } at every moment the observed direction of {the} polarization vector is determined by {the} instantaneous position of {the} vector ${\bf  e}$ projected to the celestial sphere {assigned by $(u, w)$ reference frame}. The vector ${\bf k}$ and $(u, w)$ {also} form the basis,  ${\bf k}$ is pointed to the observer at the angle $i$ to the jet axis.

Then the direction of {the} electric vector in the observer frame is determined by polarization angle $\chi$:
$$
\tan \chi = \frac{e_u}{e_w} = \frac{\cos \psi \cos i - \tan \theta_0 \sin i}{ \sin \psi},
$$
{where $\theta_0$ is the angle between ${\bf e}$ and the plane of {the} instantaneous particle position, or $(x,y)$ plane. Actually, $\theta_0$ is equal to a cone half-opening angle: $\theta_0 = \theta$. }

In \citep{Nalewajko2009} it {is} proved that polarization angle is {Lorentz invariant}, i.e. {does not transform in the observer and source reference frames, while for polarization degree (P.D.) $P$ it is not true}. In case of synchrotron radiative mechanism in {the} optical thin jet with {the} helical magnetic field{,} the dependence is implemented \citep{Lyutikov2005,Raiteri2013}:
$$
P \sim P_{max} \sin^2 \xi ',
$$
where $P_{max}$ is the highest observed P.D., $\xi '$ is the angle between {the} observer and {the} instantaneous direction of the speed $v$ in {the} source  frame. $\xi '$ depends on $\xi$ in {the} observer frame by means of Lorentz transformations:
$$
\sin \xi ' = \frac{\sin \xi}{\Gamma (1 - \beta \cos \xi)} = {\delta} {\sin \xi},
$$
where $\delta=(\Gamma(1-\beta \cos \xi))^{-1}$ is Doppler factor,  $\Gamma = (1-\beta^2)^{-1/2}$ is Lorentz factor. In {the} observer frame $\xi$ could be {determined}:
$$
\cos \xi = \frac{(\dot{\rho} \cos \psi  - \rho \dot{\psi} \sin \psi ) \sin \zeta \sin i + \dot{z} \cos \zeta \cos i}{\sqrt{\dot{\rho}^2 + {\rho}^2 \dot{\psi}^2 + \dot{z}^2}}.
$$

To calculate the model numerically for the object {S5} 0716+714 the{ jet kinematic} parameters obtained majorly by radio observations were assumed. The inclination angle $i$ and half-opening angle of the cone {$\theta_0$} were evaluated from radio observations by \citet{Pushkarev2009}: $i=5^{\circ}$, $\theta_0=1.5^{\circ}$. The value of maximum P.D. was measured {by long monitoring} \citep{Larionov2013}: $P_{max} = 0.3$. {The} physical speed of {the optical jet components was confirmed to be equal} $\beta > 0.999c$  \citep{Butuzova2017}, that {gave the} Lorentz factor $\Gamma \approx 30$. {A} pitch-angle supposed to be small {due to the model stability to its variations}. The initial distance {to the} $z$-axis was appreciated as maximum:
$$
\rho_0 = c \Delta t \approx 5 \cdot 10^{-5} \ pc,
$$
where $\Delta t \approx 1.5 $ hour is {the period} of {the} flux and polarization variations obtained from our monitoring. The {maximum} value of angular momentum $j$ was used:
$$
j_{max} = \frac{L}{E_{kin}} c = \beta \rho_0.
$$

\begin{figure}
  \centering
  \subfigure[]{\includegraphics[width=0.4\textwidth,angle=90]{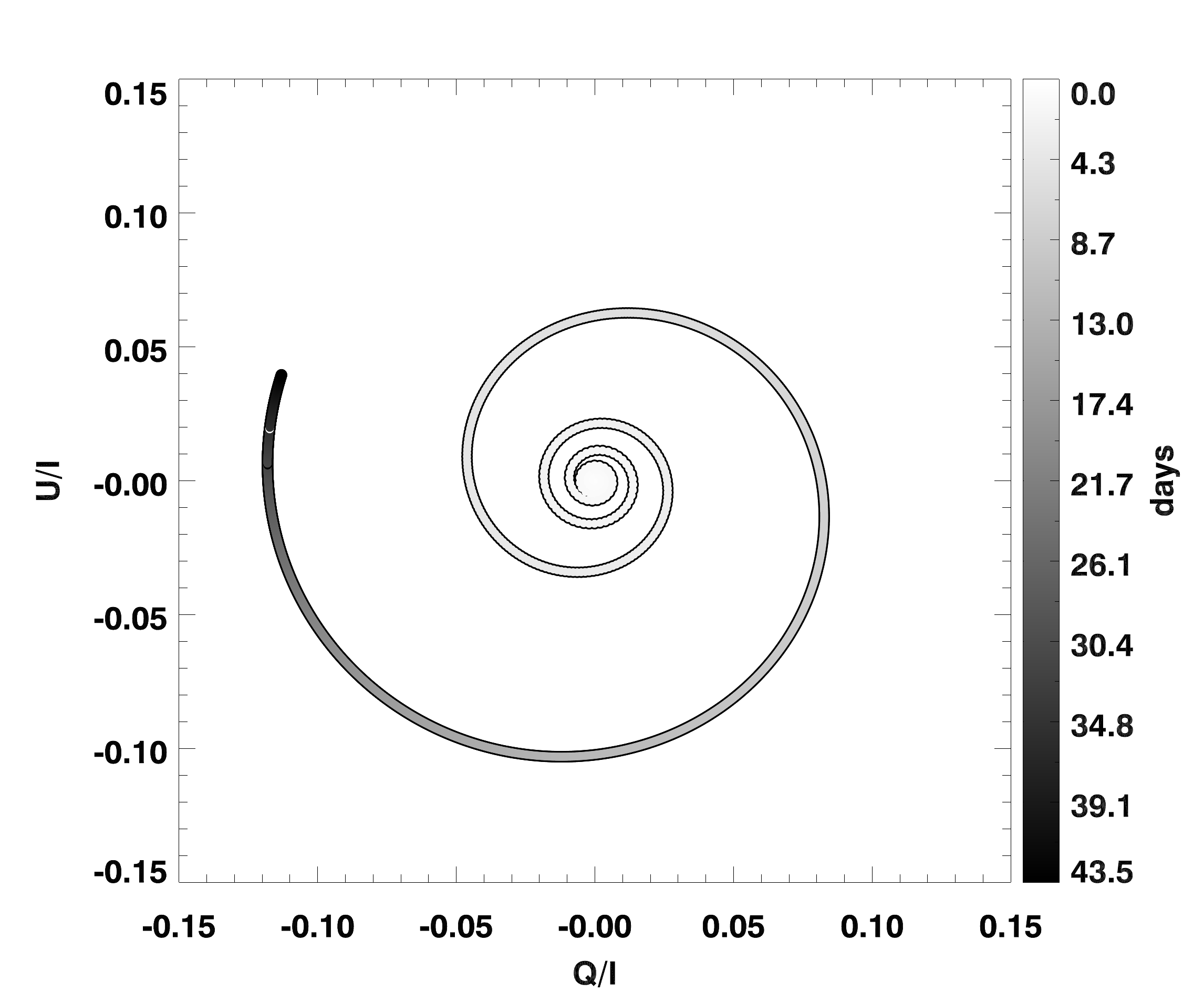}\label{QU_mod:a}}
  \subfigure[]{\includegraphics[width=0.4\textwidth,angle=90]{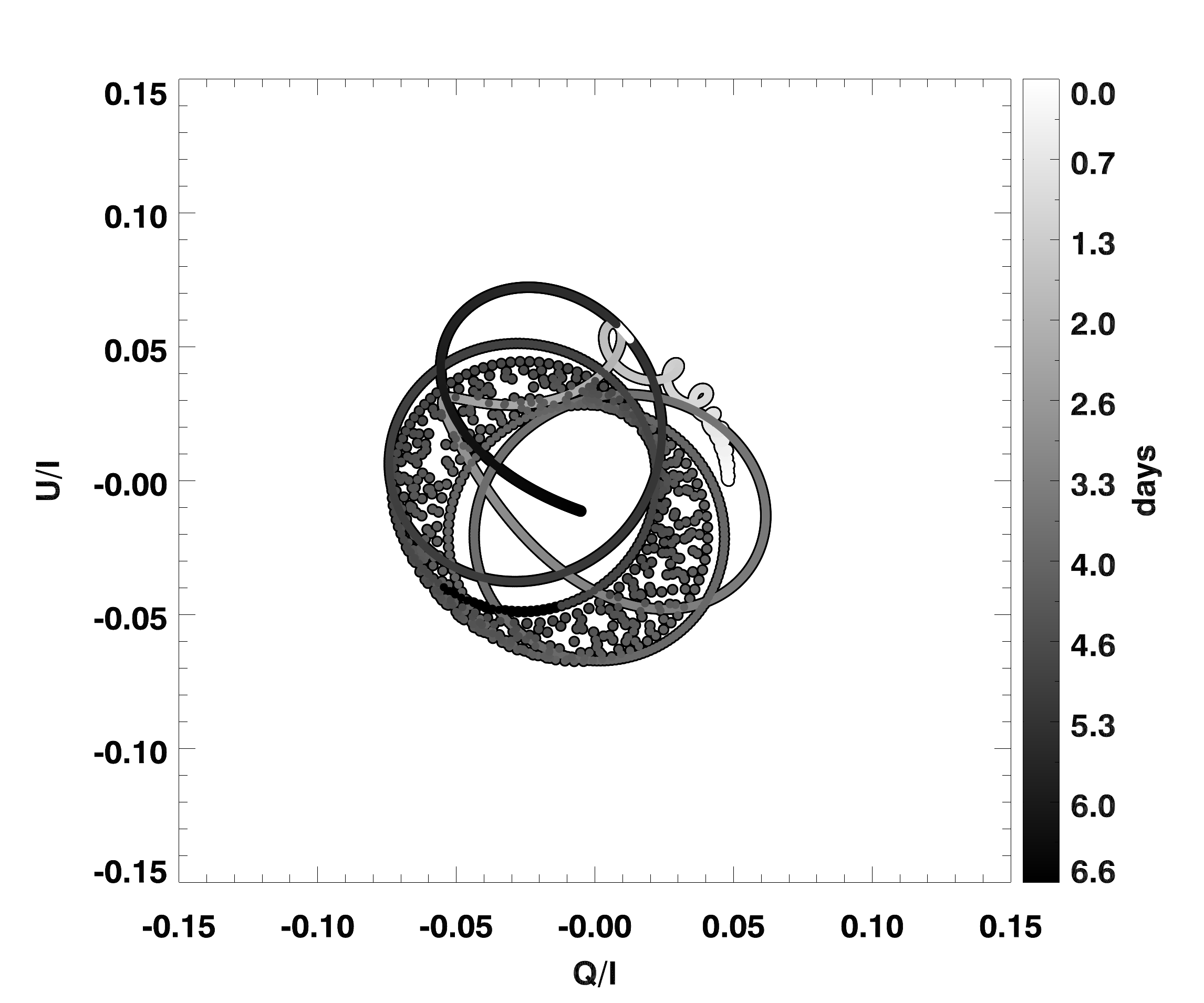}\label{QU_mod:b}}
  \caption{The results of {the} numerical model of polarization in jet: {(a) without precession and (b) with precession.}}
  \label{QU_mod:a,b}
\end{figure}

The result of {the} numerical model {processing} of {the} polarization vector {motion is demonstrated on the} $QU$-diagram. In Fig. \ref{QU_mod:a} {the} model of linear polarization variations during $\sim 40$ days is shown. The time is {marked by {colour}. The} absolute value of polarization -- P.D. -- rises rapidly within the first several days (it corresponds to the "detwisting"\ on $QU$-plane); then polarization begins to change slowly, the absolute value stays approximately constant. Eventually, the polarization vector {produces} a spiral on {the} $QU$-plane.

Therefore, {the jet geometry could cause the observed polarization vector variations on the $QU$-diagram}. Yet stable helical magnetic field structure could not be responsible for {such complex trajectory: for instance, in} \citep{Impey2000} {changing from night to night arches were recognized on the $QU$-plane but without any evidence of common spiral production. The helical motion could be assumed to be burdened with an additional component. A precession appears as the most preferable kinematic effect since it is observationally confirmed for different types of AGNs}, e.g. for Sy 1.5 Mrk 6 \citep[e.g.][]{Afanasiev2014}. Here we consider precession is one of the possible reasons for {the} observed polarization variations. 

The precession would be described as an additive value to the equations of {the} charged particle motion in {the} helical magnetic field. In this case the ${\bf E_p}$ component is added to {the} electric vector value ${\bf E}$. ${\bf E_p}$ is directed tangentially to the circle trajectory with the period of full cycle $T=2\pi/\Omega$, the absolute value is constant. Then in the celestial frame $(u,w)$ ${\bf E_p}$ would be equal to:
$$
{\bf E_p} = (\frac{\sin \Omega}{\cos i},\cos \Omega),
$$
consequentially, the electric vector position angle {is} determined:
$$
\tan \chi_p = \frac{\sin \Omega}{\cos i \cos \Omega}.
$$

To calculate the model with precession numerically we used parameters fitting best to our observational data. The selected parameters are presented in Tab. \ref{t4}, the result of {modelling could be found in} Fig. \ref{QU_mod:b}. The model is significantly sensitive to {the} precession period $T$. It should be clarified that the observed {precession} motion of several {active galaxies jets} concerns the {radio jet scales}  {and} average periods of years. Since we investigate much more compact regions  --  $10^{-3} $ pc, the much less period should be considered. Varying the model parameters we would like to fit {average polarization variations particularly} -- both the form and {the} switch time. Finally, we { have found} out that the observed significant distortion of the particle motion could be explained only in case of small time-scales of {the} additive kinematic effect.

\begin{figure}
\centering
\includegraphics[width=0.85\columnwidth, angle=90]{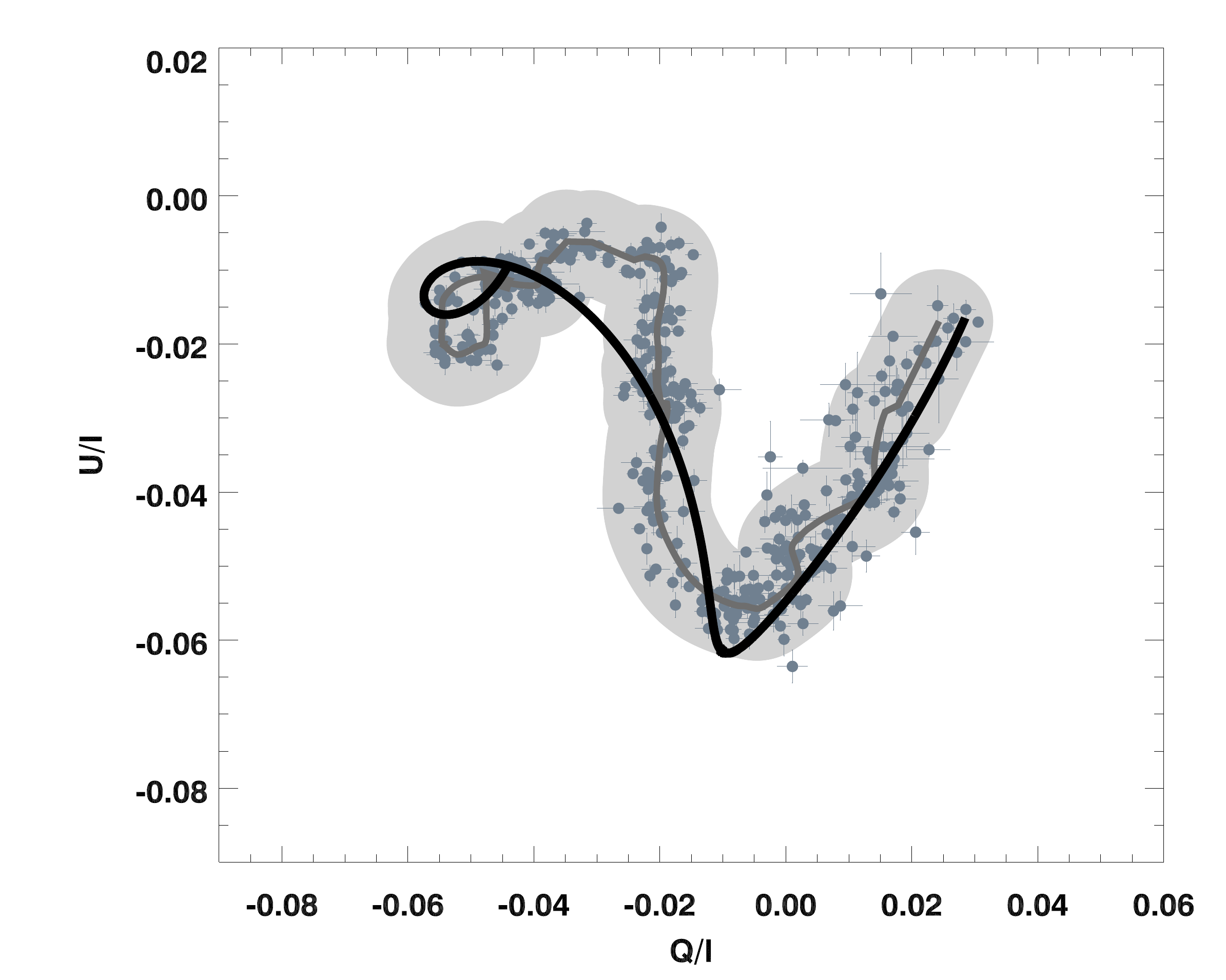}
\caption{The model of linear polarization on {the} $QU$-plane in case of {jet precession}. The observational data {are plotted with grey dots with error bars. The $3\sigma$ confidence area is {coloured} light grey.}}
\label{QU_90}
\end{figure}

\begin{table}
  \begin{center}
    \caption{The parameters of numerical model of  {S5} 0716+714 polarization with confirmed precession.}
    \label{t4}
    \begin{tabular}{l c c}  
    
    \hline
    {Parameter} &   & Value \\
     \hline
        Distance from the precession axis & $\omega$ & $ 0.7 \cdot 10^{-3} $ pc \\
        Precession period & $T$ &  15 days \\
        
     \hline
    \end{tabular}
  \end{center}
\end{table}

In Fig. \ref{QU_90} we present the result of fitting our data by the model of polarization in {the} helical magnetic field precessing with the period $\sim 15$ days. {To testify to the goodness of our fit we have added a  $3\sigma$ confidence area of the smoothed polarization vector rotation. One could discover that the model with the parameters described above assuredly fit the data with respect to the measurement errors. The exception is an interval of 21-22 hours, where the model curve falls out of the estimated area, yet we rely on the physical processes not considered in our geometrical model while it actually does not affect our quantitative outcomes.}

\section{Discussion} \label{disc}

The phenomenon of the {S5 0716+714 intraday variability} was suspected for a long time. However, {to reveal the variations on the time-scale of hours the high accuracy and time resolution had been needed, that was not performed in long-term variability investigations. While} the observations {carried out} by our group {have met} these demands. 

One of the first investigations devoted to {revealing of} {S5} 0716+714 rapid variability was the work \citep{Impey2000}, where the monitoring of the object {at} the BTA in 1991 and 1994 was presented. The data plotted on $QU$-plane circumscribed the {apparent} arches changing their positions from night to night. Due to the short sets (less {than} $2$ hours){,} the unambiguous conclusions on the origin of changing electric vector direction could not be done. 

In \citep{Bhatta2015} the data of 16-hour polarimetric monitoring of {S5} 0716+714 was given. Comparing with \citep{Impey2000} these observations have an order less time resolution -- $\sim 20$ min and {the} accuracy of polarization measurement 2-10\%. Though, the observation set was longer that {discovered} an approximately closed arch on {the} $QU$-diagram.

One more investigation of {the} {S5} 0716+714 rapid variability was {declared} in \citep{Sasada2008}. It is {essential} that the data obtained by authors also {perform} an arch on {the} $QU$-plane. Moreover, in this paper{,} the average time of flux variations was calculated as $\sim 15$ min. There the average time meant only the rising time of the bump, that {agreed} with the variability time-scale of $\sim 1.5$ hours, {discovered} by our observations, as we {took} into account not only the time of flux rising but the whole cycle of flux change. As a result, the assertion done in the paper that the size of the emitting region {could be} defined by 15-minutes fluctuations {seemed} incorrect.

\begin{figure}
    \centering
        \includegraphics[width=0.83\columnwidth, angle=90]{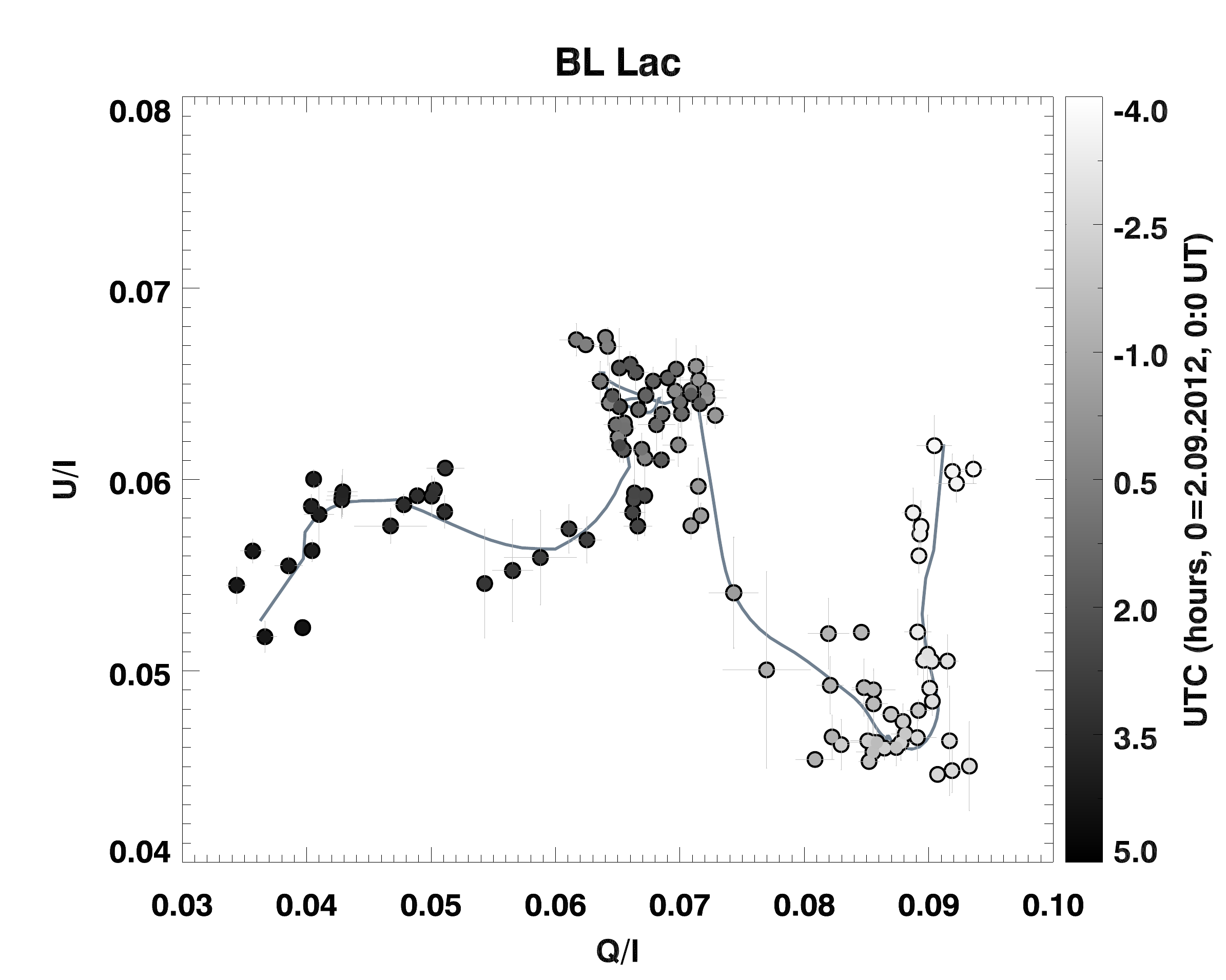}
    \caption{The variations of the Stokes parameters $Q$ and $U$ of BL Lac within 9 hours \citep{Covino2015}.}
    \label{bllac}
\end{figure}

It should be mentioned that the variations of polarization vector along the arches on {the} $QU$-plane {has been} registered {at least} for one more blazar -- BL Lac. In \citep{Covino2015} the 9-hour polarimetric monitoring of BL Lac was provided with the almost 1 min resolution and high accuracy of polarization measurement -- $\sim 0.2-0.3 \%$. The monitoring was intermittently interrupted to observe the standard outside the field, but, nevertheless, the long {dataset} allowed to draw conclusions of variability pattern within the night. In the paper the polarization degree and angle {were originally} presented, yet we redrew the data on the $QU$-plane (Fig. \ref{bllac}) revealing that {the} polarization vector {were changing} along the trajectory similar to {S5} 0716+714 {one}, and the time of the direction switch {was} close to the time calculated from our observations.

According to 9-hour monitoring{,} we are able to claim the intraday variability of flux and polarization of {S5} 0716+714. As $QU$-plane reflects directly the plasma motion in the jet, the observed "arches" and "loops" bear out the common suggestions of {the} helical configuration of {the} magnetic field. Indeed, {the ordered magnetic field with {the} helical structure in common produces the "arch"-like} trajectories of charged particles {radiating} non-thermal polarized emission. Yet {as it was shown above} in case of absence of extra kinematic effects the polarization {moves} along the rings. Much {more} complicated paths on {the} $QU$-plane could be explained with additional precession motion {of} either {the} magnetic field inside the jet or the outflow itself due to the precession of {the} central black hole. In the last case{,} the {period agreed with observations} seems too {short, so} the mechanism of such rapid jet "walk" is not clear. On the contrary, due to magnetohydrodynamics{,} the precession of {the} magnetic field on short time-scales is founded better.

To sum up, the observed phenomenon of rapid polarization variations could be qualitatively clarified only with geometrical effects. It is obvious that the {submitted} model is not absolutely accurate: it depends on a variety of parameters, not obtained for {the} optical jet, and also does not {take into account} the physical processes in plasma. {Though,} it could be distinctly declared that the geometrical effects dominate and highly accurately explain the {variability}. 

{Moreover, the} time of polarization vector switch is very similar to the period of flux variations: {$\sim 80$ min for polarimetry and at least $1.5$ hours for photometry. Comparing figures} \ref{wv} and \ref{dragon} {one could see that even the dynamics of the variability development is the same: wavelet transform becomes inefficient when polarization begins to move along the arch without switching. Such behaviour} adjusts with {the earlier presented geometrical models} \citep{Butuzova2017}: { while the emitting matter moves along the jet the Doppler factor varies that influences the boost of the relativistic plasma radiation. Therefore, the plasma motion in the jet affects not only the polarization vector rotation but the flux intensity as well that is in good agreement with the observations discovering the similarity of the total flux and polarization variations time-scale.} 

The essential profit of our model of polarization in {the} jet is that {as the polarization variation time -- switch time -- and period of flux variations are equivalent} they are closely connected to the linear scales of {the} emitting region. According to the common jet model \citep{Gabuzda2013}, the plasma moves at the sides of the jet cone and, therefore, the time of polarization making the "loop" on {the} $QU$-plane {could be used to derive} the size of the area: $r=c \Delta t$ and, in case of our {results}, $r \approx 1.5$ light hour or $5 \cdot 10^{-5}$ pc. This result is particularly important as the optical jet could not be resolved with modern techniques, and our estimation is the only possible indirect measurement of emitting area size.

\section{Conclusions}

In {the} given paper the intraday variability of {S5} 0716+714 was investigated and the following was done:

\begin{itemize}
    \item[(i)] during the 9-hour polarimetric monitoring carried out by the 6-m telescope the intraday variability of {S5} 0716+714 was revealed in both total (with an amplitude 0.04 mag) and polarized ($\Delta P.D. = 7\%$) light on the time-scale $\sim 1.5$ hour;
    \item[(ii)] using $QU$-plane to {demonstrate the} polarization vector variations we discovered the pattern of changes in polarized flux: within the night {the} polarization vector circumscribes "arches" and "loops". The similar scene was found for {S5} 0716+714 in several preceding papers \citep[e.g.][]{Impey2000, Bhatta2015}, but the time resolution and accuracy were not valid enough. Such {behaviour} could not be revealed by plotting P.D. and polarization angle separately {because of the vector quantity changes};
    \item[(iii)] according to the general model of jet structure \citep{Gabuzda2013}, the time of polarization vector variations is directly connected to the linear size of the emitting region: 1.5 light hour or $5 \cdot 10^{-5}$ pc at the $\sim 10^{-3}$ pc distance from the central black hole (optical jet). As optical jet is unresolved by modern techniques, we offered the results of indirect measurements based on polarimetric observations that {are} especially essential for jet physics researches;
    \item[(iv)] we suggested a model of polarization producing by geometrical effects due to relativistic plasma motion in {the} helical magnetic field and, moreover, we added a feature, {particularly new in numerical modelling}, that only could qualitatively explain rapid polarization variability -- the precession of {the} helical magnetic field; 
    \item[(v)] our model of relativistic plasma motion in {the} precessing magnetic field qualitatively and numerically describes our observational data; {by} fitting {the} model to the data{,} we calculated the period of precession -- approximately 15 days. Furthermore, our model is able to reconduct the pattern of polarization vector motion on $QU$-plane for BL Lac monitoring \citep{Covino2015} as well.
\end{itemize}

\section*{Acknowledgements}

We sincerely thank V.R. Amirkhanyan for valuable discussions and useful remarks. The results of observations were obtained with the 6-m BTA telescope of the Special Astrophysical Observatory Academy of Sciences, operating with the financial support of the Ministry of Education and Science of Russian Federation.  

%%%%%%%%%%%%%%%%%%%%%%%%%%%%%%%%%%%%%%%%%%%%%%%%%%



%\bibliographystyle{mnras}


%%%%%%%%%%%%%%%%%%%%%%%%%%%%%%%%%%%%%%%%%%%%%%%%%%

%%%%%%%%%%%%%%%%% APPENDICES %%%%%%%%%%%%%%%%%%%%%

%\appendix


%%%%%%%%%%%%%%%%%%%%%%%%%%%%%%%%%%%%%%%%%%%%%%%%%%

% Don't change these lines
\bsp	% typesetting comment
\label{lastpage}
\end{document}